\documentclass[twocolumn,prl,aps,superscriptaddress,showpacs]{revtex4-1}

\usepackage{amsmath}
\usepackage{amssymb}
\usepackage{bbm}
\usepackage{braket}
\allowdisplaybreaks
\usepackage{pdfpages}
\usepackage{graphicx}
\usepackage[colorlinks=true,citecolor=blue,linkcolor=blue]{hyperref}
\usepackage{color}                                      
\definecolor{d_red}{cmyk}{0.00, 0.81, 1.00, 0.27}
\definecolor{d_orange}{cmyk}{0.00, 0.33, 1.00, 0.00}
\definecolor{d_blue}{cmyk}{0.78, 0.47, 0.00, 0.20}
\definecolor{d_lgreen}{cmyk}{0.07, 0.00, 0.79, 0.29}
\definecolor{d_green}{cmyk}{0.66, 0.00, 0.71, 0.56}
\definecolor{d_blue}{cmyk}{0.78, 0.47, 0.00, 0.20}
\definecolor{d_dblue}{cmyk}{0.91, 0.79, 0.00, 0.22}
\definecolor{d_pink}{cmyk}{0.0, 0.79, 0.37, 0.29}
\definecolor{d_purple}{cmyk}{0.16, 0.54, 0.00, 0.70}
\definecolor{d_paleblue}{cmyk}{0.669, 0.338, 0.00, 0.373}
\definecolor{d_dpaleblue}{cmyk}{0.441, 0.290, 0.00, 0.580}
\definecolor{d_brown}{cmyk}{0.0, 0.490, 0.930, 0.350}
\definecolor{d_turquoise}{cmyk}{0.630, 0.04, 0.0, 0.440}

\newcommand{\av}[1]{\langle #1 \rangle}

\def\bmx{\begin{pmatrix}}
\def\emx{\end{pmatrix}}

\usepackage{hyperref}
\usepackage[figure,table]{hypcap}               
\hypersetup{
pdftitle = {},
pdfsubject = {},
pdfauthor = {},
pdfkeywords = {},
pdfcreator = {},
pdfproducer = {LaTeX with hyperref},
colorlinks = true,
linkcolor = blue, 
anchorcolor = blue,
citecolor = blue, 
filecolor = red, 
menucolor = red, 
pagecolor = red, 
urlcolor  = blue, 
breaklinks = true,
pdfstartview = FitV,
pdfhighlight = /I,
pdfpagelayout = OneColumn,
hypertexnames=true 
}

\allowdisplaybreaks

\usepackage{graphicx}

\begin{document}

\title{Universal post-quench prethermalization at a quantum critical point}

\author{Pia Gagel}
\affiliation{Institute for Theory of Condensed Matter, Karlsruhe Institute of
Technology (KIT), 76131 Karlsruhe, Germany}
\author{Peter P. Orth}
\affiliation{Institute for Theory of Condensed Matter, Karlsruhe Institute of
Technology (KIT), 76131 Karlsruhe, Germany}
\author{J\"org Schmalian}
\affiliation{Institute for Theory of Condensed Matter, Karlsruhe Institute of
Technology (KIT), 76131 Karlsruhe, Germany}
\affiliation{Institute for Solid State Physics, Karlsruhe Institute of Technology
(KIT), 76021 Karlsruhe, Germany}


\date{\today }
\begin{abstract}
We consider an open system near a quantum critical point that is suddenly
moved towards the critical point. The bath-dominated diffusive non-equilibrium
dynamics after the quench is shown to follow scaling behavior, governed
by a critical exponent that emerges in addition to the known equilibrium
critical exponents. We determine this exponent and show that it describes
universal prethermalized coarsening dynamics of the order parameter
in an intermediate time regime. Implications of this quantum critical
prethermalization are a powerlaw rise of order and correlations after
an initial collapse of the equilibrium state and a crossover to thermalization
that occurs arbitrarily late for sufficiently shallow quenches. 
\end{abstract}
\maketitle
Predicting the out-of-equilibrium dynamics of quantum many-body systems
is a challenge of fundamental and practical importance. This research
area has been boosted by recent experiments in cold-atom gases~\cite{bloch:885}
and scaled-up quantum-circuits~\cite{HouchTuereciKoch-NatPhys-2012},
by ultra-fast pump-probe measurements in correlated materials~\cite{Fausti14012011,Smallwood01062012,LiWang-FemtosecondSwitchingMagnetism-Nature-2013},
and by performing heavy-ion collisions that explore the quark-gluon
plasma~\cite{Arsene20051}. In this context, the universality near
a quantum critical point (QCP), well established in and near equilibrium,
comes with the potential to make quantitative predictions for strongly
interacting systems far from equilibrium. For example, the quantum
version~\cite{PhysRevLett.95.035701,0295-5075-84-6-67008,PhysRevB.84.224303,PhysRevB.86.064304,PhysRevLett.109.015701}
of the Kibble-Zurek mechanism of defect formation~\cite{Kibble-JPhysA-1976,Zurek-Nature-1985}
was developed for systems driven through a symmetry breaking QCP at
a small, but finite rate. Similarly, near a QCP the long-time dynamics
after a sudden change of Hamiltonian parameters, is governed by equilibrium
exponents~\cite{PhysRevB.81.012303}. These phenomena occur in the
regime of longest time scales.

Recently, however, many physical systems away from equilibrium were
identified which display novel dynamical behavior on intermediate
time scales, a behavior often referred to as prethermalization~\cite{PhysRevLett.93.142002,PhysRevB.75.144418,PhysRevLett.103.056403,PhysRevLett.100.175702,1367-2630-12-5-055008,PhysRevB.87.205109,PhysRevLett.106.050405,PhysRevLett.98.180601,PhysRevLett.98.210405,PhysRevLett.111.197203,PhysRevLett.110.136404}.
The question arises whether one can expect universality during prethermalization
if one drives a system towards a QCP. Even if this is done at a finite
rate $1/\tau$, a system will fall out of equilibrium at some point,
a behavior owed to the critical slowing down near the QCP. Then a
scaling theory with characteristic time scale $\tau$ can be developed~\cite{PhysRevB.86.064304},
where regions of the size of the freeze-out length $\propto\tau^{1/z}$
emerge that behave like in equilibrium. $z$ is the dynamic critical
exponent. In case of a quantum quench, the time scale $\tau$ and
the freeze-out length become comparable to microscopic time and length
scales, respectively and the system instantly falls out of equilibrium.
The detailed recovery of this out-of-equilibrium dynamics, along with
the time dependence of length scales, order-parameter correlations,
and the potential for out-of-equilibrium universality are major theoretical
and experimental challenges. 
\begin{figure}[b!]
   \centering
   \includegraphics[width=.85\linewidth]{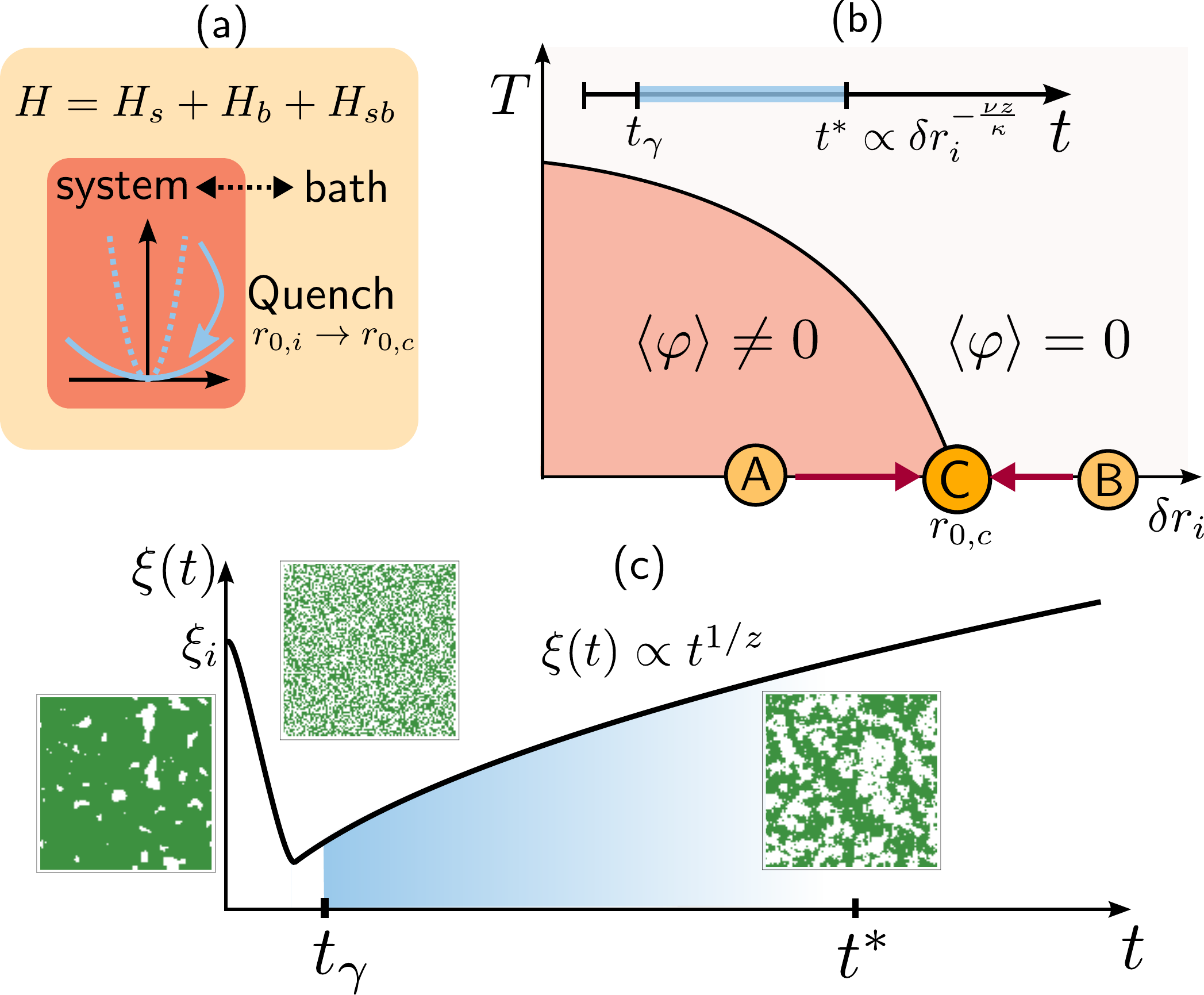}
   \caption{(Color online) (a) Schematic description of the setup and quench protocol. 
(b) Schematic phase diagram as a function of temperature $T$ and mass $\delta r_{i} = r_{0,i} - r_{0,c}$. 
Red arrows describe the quench protocol. Dynamics exhibits
three time regimes: $t < t_\gamma=\gamma^{-z/2(z-1)}$ with non-universal dynamics, the universal prethermalized regime $t_\gamma < t < t^* \propto \delta r_i^{-\nu z/\kappa}$ that we study, and a quasi-adiabatic regime $t > t^*$ described by equilibrium critical exponents. Here, $\gamma$ is system-bath coupling and $\kappa/\nu$ the scaling dimension of $\delta r_i$. (c) Correlation length collapse and light-cone like revival following a quench with initial length $\xi_i$. Inset: sketches of order parameter configurations with domains of typical size $\xi(t)$. }
   \label{fig:1}
 \end{figure}

In this Letter, we show that the time evolution of observables in
an open system that is suddenly moved to a QCP displays universal
behavior (see Fig.~\ref{fig:1}(a-b)). Their non-equilibrium dynamics is governed by a critical exponent that describes the slow decay of post-quench correlations and response soon after a quantum-quench, where initial correlations are still important. It is therefore not
related to equilibrium exponents. This behavior is astounding as
universality is usually reserved for large time and length scales.
From the value of the exponent we conclude that initial state correlations
rapidly collapse after a quench and that the order parameter undergoes
an intermediate coarsening, i.e. grows due to the growing \emph{light-cone}
length $\xi\left(t\right)\propto t^{1/z}$ (see Fig.~\ref{fig:1}(c)), before it decays quasi-adiabatically
at longer times. We also demonstrate that the duration of this intermediate
prethermalization can be manipulated and tuned to be arbitrarily large.
While there are important differences between classical and quantum
quenches, the analysis of this paper was motivated by the pioneering
theory of classical dynamics in Ref.~\onlinecite{JanssenSchaubSchmittmann-1989} (see also Ref.~\cite{1742-5468-2012-01-P01014} for the case of colored noise). 

The quench protocol that underlies our analysis is indicated in Fig.~\ref{fig:1}(a-b).
We consider a quantum many-body system that is coupled to an external
bath of harmonic oscillators. Prior to the quench, the complete system
is prepared in the ground state of the initial Hamiltonian $H_{i}=H_{s,i}+H_{b}+H_{sb}$.
The initial Hamiltonian of the system, $H_{s,i}$, describes a $N$-component
scalar quantum field $\mathbf{\boldsymbol{\varphi}}\left(x,t\right)$
with components $\varphi_{a}$ ($a=1,\ldots,N$): 
\begin{equation}
H_{s,i}=\frac{1}{2}\int d^{d}x\left(\boldsymbol{\pi}^{2}+r_{0,i}\boldsymbol{\varphi}^{2}+\left(\nabla\boldsymbol{\varphi}\right)^{2}+\frac{u}{2}\boldsymbol{\varphi}^{4}\right)\,,\label{eq:Hamilt}
\end{equation}
where $\boldsymbol{\pi}$ is the canonically conjugated momentum to
$\mathsf{\boldsymbol{\varphi}}$. $H_{b}=\frac{1}{2}\int d^{d}x\sum_{l}\left(\Omega_{l}^{2}\boldsymbol{X}_{l}^{2}+\boldsymbol{P}_{l}^{2}\right)$
describes the external bath of of harmonic oscillators and $H_{sb}=\sum_{l}c_{l}\int d^{d}x\boldsymbol{X}_{l}\cdot\boldsymbol{\varphi}$
the coupling between system and bath. Next, we suddenly change $H_{s,i}\rightarrow H_{s}$
by switching $r_{0,i}\rightarrow r_{0,c}$ to its value right at the
QCP of system + bath in equilibrium (see Fig.~\ref{fig:1}(b)). The
time evolution after the quench is now governed by the new Hamiltonian
$H=H_{s}+H_{b}+H_{sb}$. The bath ensures that the system eventually
equilibrates at $T=0$, which allows reaching the QCP for $t\rightarrow\infty$. 
A crucial variable is the distance to the critical point  $\delta r_{i}=r_{0,i}-r_{0,c}$
before the quench. After the quench we consider $\delta r_{f}=r_{0,f}-r_{0,c}=0$,
while the same behavior is expected for generic quenches that move
the system closer to the critical point $\delta r_{f}\ll\delta r_{i}$
or take place at a finite but small temperature $T\ll\delta r_{i}^{\nu z}/\gamma^{z/2}$ with system-bath coupling $\gamma$ defined below. 

The Hamiltonian $H_{s,i}$ of Eq.\eqref{eq:Hamilt} describes a transverse-field
Ising model for $N=1$, systems near a superconducting-insulator quantum
phase transition, Josephson junction arrays and quantum antiferromagnets
in an external magnetic field for $N=2$, or quantum dimer systems
for $N=3$~\cite{sachdev_qpt_book}. Our theory for system + bath
can then be applied to a range of systems such as dissipative superconducting
nano-wires~\cite{sachdev_universal_2004}, the superfluid-insulator
transition in cold-atom gases coupled to other bath-atoms~\cite{PhysRevLett.105.045303},
or low-dimensional Heisenberg spin-dimers or transverse field Ising
spins with strong quantum fluctuations and coupling to phonons. Another promising realization can be achieved by an ensemble of qubits in a photon cavity~\cite{HouchTuereciKoch-NatPhys-2012,koch:023811}. The
effects of the bath are described in terms of $\eta\left(\omega\right)=-\sum_{l}\frac{c_{l}^{2}}{\left(\omega+i0^{+}\right)^{2}-\Omega_{l}^{2}}$.
We consider for the spectral density of the bath: 
\begin{align}
\mathrm{Im}\,\eta\left(\omega\right)=\gamma\omega\left|\omega\right|^{\alpha-1}e^{-|\omega|/\omega_{c}}\label{eq:5}
\end{align}
with damping coefficient $\gamma$ and cut-off energy $\omega_{c}$.
The exponent $\alpha$ determines the low-energy spectrum of the bath,
where $\alpha=1$ corresponds ohmic damping while $\alpha>(<)1$ corresponds
to super-ohmic (sub-ohmic) damping~\cite{weissdissipation}. In the
following, we consider the hierarchy of scales $\omega_{c}\gg t_{\gamma}^{-1}=\gamma^{1/(2-\alpha)}$
and analyze the regime $t>t_{\gamma}$ when the dynamics is dominated
by the bath. For the one-loop RG analysis used in this paper, it
holds $z=2/\alpha$.

\begin{figure}[t!]
   \centering
   \includegraphics[width=.65\linewidth]{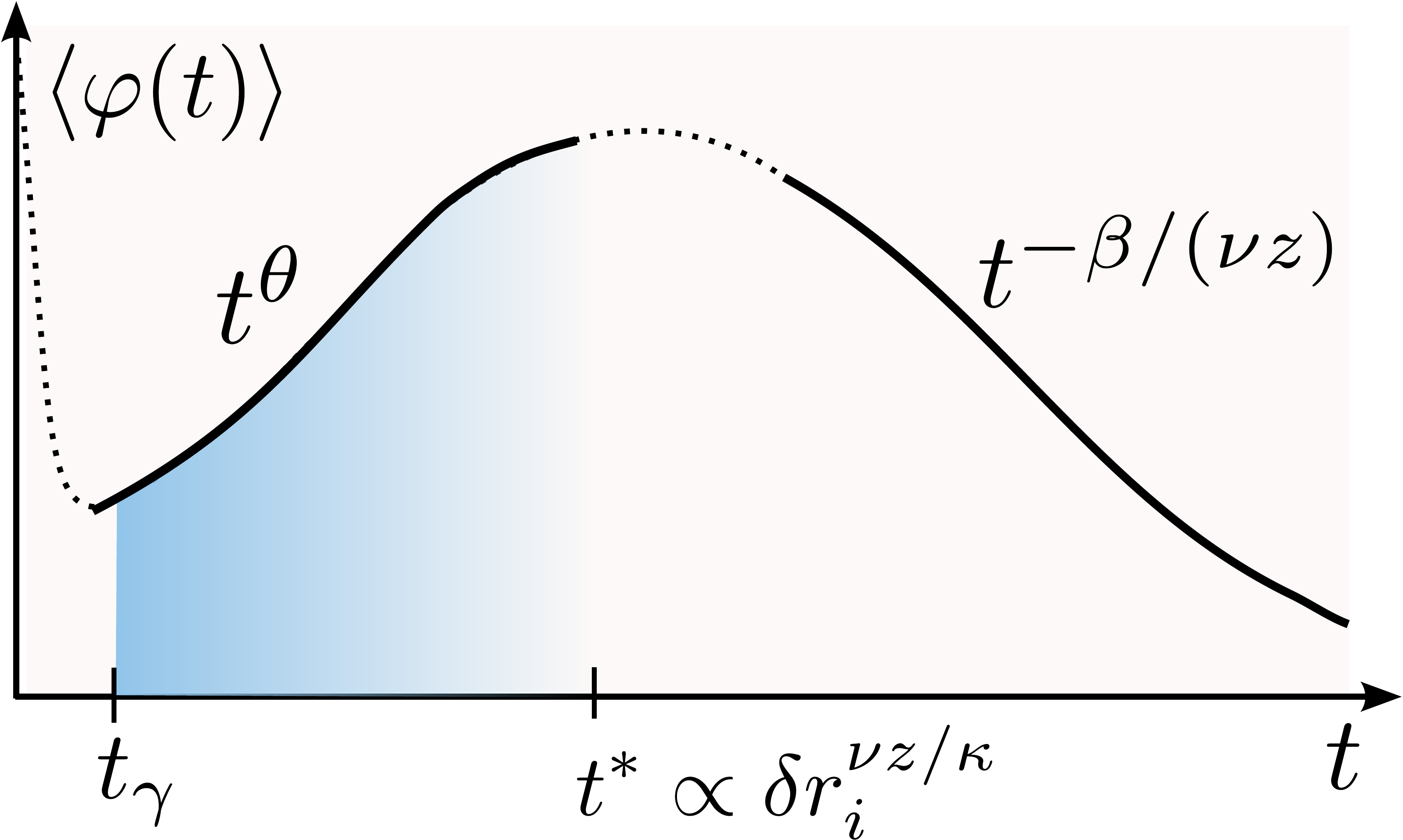}
   \caption{(Color online) Schematic order parameter dynamics $\av{\varphi(t)}$. In the prethermalized regime $t_\gamma< t < t^*$ (blue) it is governed by new universal critical exponent $\theta$. At longer times, $\av{\varphi(t)}$ decays to zero quasi-adiabatically as described by equilibrium exponents.} 
   \label{fig:2}
 \end{figure}

We start with general scaling arguments for the non-equilibrium dynamics
after a quench towards the QCP. The scaling behavior will be confirmed
using a perturbative renormalization group (RG) analysis later in
the paper. In equilibrium, the order parameter behaves as function
of the distance $\delta r$ to the QCP according to $\left\langle \varphi_{a}\left(\delta r\right)\right\rangle _{{\rm eq}}=b^{-\beta/\nu}\left\langle \varphi_{a}\left(b^{1/\nu}\delta r\right)\right\rangle _{{\rm eq.}}$
, with scaling parameter $b>1$, which leads to the well known behavior
$\left\langle \varphi_{a}\left(\delta r\right)\right\rangle _{{\rm eq}}\propto\delta r^{\beta}$.
In our case $\delta r$ rapidly changes as function of time from $\delta r_{i}$
to $\delta r_{f}$, leading to a $t$-dependence of the order parameter.
The generalization of the equilibrium scaling relation can be performed
in analogy to boundary layer scaling theory as it occurs near surfaces
and interfaces~\cite{doi:10.1142/S0217979297001751}. Here, a new
healing length scale associated with surface fields appears.
In our problem, the boundary layer incorporating the initial value
problem corresponds to a ``surface in time''\cite{PhysRevLett.96.136801,CalabreseCardy-JStatMech-2007,0295-5075-95-6-66007}
and exhibits an associated new healing time scale $t^{*}$.
Following Ref.~\onlinecite{doi:10.1142/S0217979297001751} it follows for
the order parameter $\left\langle \varphi_{a}\left(\delta r_{i},\delta r_{f},t\right)\right\rangle _{{\rm }}=b^{-\beta/\nu}\left\langle \varphi_{a}\left(b^{\kappa/\nu}\delta r_{i},b^{1/\nu}\delta r_{f},b^{-z}t\right)\right\rangle $.
While $\delta r_{f}$ scales as in equilibrium, reflecting the fact
that the system approaches equilibrium for $t\rightarrow\infty$,
the initial mass $\delta r_{i}$ has a nontrivial scaling exponent
$\kappa/\nu$. For $\delta r_{f}=0$, i.e. a quench right to the QCP,
it follows with $b=t^{1/z}$: 
\begin{equation}
\left\langle \varphi_{a}\left(t,\delta r_{i}\right)\right\rangle =t^{-\frac{\beta}{\nu z}}\Phi\left(t^{\frac{\kappa}{\nu z}}\delta r_{i}\right),\label{eq:scall-1}
\end{equation}
with universal function $\Phi\left(y\right)$. As shown in Fig.~\ref{fig:2}, in the long time limit $t\gg t^{*}$ for $\Phi\left(y\gg1\right)\rightarrow\text{const}.$, the order parameter decays quasi-adiabatically ( $\left\langle \varphi_{a}\left(t\right)\right\rangle \propto t^{-\frac{\beta}{\nu z}}\propto\xi\left(t\right)^{-\beta/\nu}$) to zero with timescale 
\begin{equation}
t^{*}\propto\delta r_{i}^{-\frac{\nu z}{^{\kappa}}}\,.
\end{equation}
In the opposite limit $t\ll t^{*}$ the situation is qualitatively
different. Assuming in analogy to Ref.~\onlinecite{doi:10.1142/S0217979297001751}
that the susceptibility with respect to a temporal boundary-layer
term $\left\langle \varphi_{a,i}\right\rangle \propto\delta r_{i}^{\beta}$$ $
is finite, it follows $\Phi\left(y\ll1\right)\propto y^{\beta}$,
such that 
\begin{equation}
\left\langle \varphi_{a}\left(t\right)\right\rangle \propto t^{\theta}\:\:\:\mathrm{with}\:\:\:\theta=\frac{\left(\kappa-1\right)\beta}{\nu z}.\label{eq:phi_of_t}
\end{equation}
Thus, a new universal time dependence of the order parameter emerges
at short times. The value of the exponent $\theta$ is determined
by the scaling dimension $\kappa$ of $\delta r_{i}.$ The time scale
$t^{*}$ separates the regime governed by the initial quench and concomitant
fall out of equilibrium from the quasi-adiabatic long time behavior.
Thus, in analogy to spatial boundary layer problems it describes the
dynamic healing after the quench.

The same exponent $\theta$ also determines the time dependence of
correlation and response functions. To analyze the non-equilibrium
dynamics we employ the Keldysh formalism of many-body theory~\cite{Kamenev-NonEqFieldTheory-Book} and use the specific form of the Keldysh contour of Ref.~\cite{Danielewicz-AnnPhys-1984}, appropriate for our quench protocol. The key quantities are the retarded response function $G^{R}$ and the Keldysh correlation function $G^{K}$:
\begin{eqnarray}
G^{R}\left(k,t,t'\right) & = & -i\theta\left(t-t'\right)\left\langle \left[\varphi_{a}\left(k,t\right),\varphi_{a}\left(-k,t'\right)\right]_{-}\right\rangle \nonumber \\
G^{K}\left(k,t,t'\right) & = & -i\left\langle \left[\varphi_{a}\left(k,t\right),\varphi_{a}\left(-k,t'\right)\right]_{+}\right\rangle
\end{eqnarray}
with momentum $k$. They are no longer related by the fluctuation-dissipation theorem.
We expect from dimensional arguments 
\begin{equation}
i G^{R(K)}\left(k,t,t'\right)=\left(\frac{t}{t'}\right)^{\theta(\theta')}\frac{f^{R(K)}\left(k^{z}t/\gamma^{z/2},t'/t\right)}{k^{2-\eta-z}\gamma^{z/2}}\,.\label{eq:scaling}
\end{equation}
In an out-of-equilibrium state the correlation and response functions
depend on both time variables. This gives rise to an additional dimensionless
ratio $t/t'$ compared to scaling in equilibrium. The singular dependence
on this ratio in $G^{R}$ and $G^{K}$ is characterized by exponents
$\theta$ and $\theta'$, respectively. Thus, the scaling functions
$f^{R}$ and $f^{K}$ depend only weakly on $t'/t$ if $t\gg t'$.
The exponents $\theta$ and $\theta'$ are not independent. Relating $G^{R}$ and $G^{K}$ in the Dyson equation yields $\theta=\theta'+\frac{2-z-\eta}{z}$.

Let us now demonstrate that $\theta$ in Eqs.~\eqref{eq:phi_of_t}
and~\eqref{eq:scaling} is indeed the same. We consider an initial
state characterized by a finite order parameter $\left\langle \boldsymbol{\varphi}_{i}\right\rangle $
(path $A\rightarrow C$ in Fig.~\ref{fig:1}(b)). A region of volume $\xi\left(t\right)^{d}$
is correlated at time $t$ after the quench and Eq.~\eqref{eq:scaling}
yields for the local, i.e. momentum averaged, correlation function
$G_{loc}^{K}\left(t,t^{\prime}\right)\propto\left(t/t^{\prime}\right)^{\theta'}t^{-\frac{d-\eta-z+2}{z}}$.
The initial order parameter $\left\langle \boldsymbol{\varphi}_{i}\right\rangle $
polarizes the system for a certain time. The magnetization at time
$t$ is then $\left\langle \boldsymbol{\varphi}_{i}\right\rangle $
multiplied by the local correlation function up to $t$ and the size
of the correlation volume: $\left\langle \boldsymbol{\varphi}\left(t\right)\right\rangle \simeq\left\langle \boldsymbol{\varphi}_{i}\right\rangle iG_{loc}^{K}\left(t,t^{\prime}\right)\times t^{d/z}$.
We obtain the powerlaw behavior of the order parameter of Eq.\eqref{eq:phi_of_t}.
The time dependence of the order parameter is therefore a balance
between the decay of local correlations encoded in $G_{loc}^{K}\left(t,t^{\prime}\right)$
and the growth of the volume encompassed by light-cone propagation,
i.e. $\xi\left(t\right)$. 

Next we demonstrate this behavior in an explicit analysis and determine
the value of the exponent $\theta$. We start using simple perturbation
theory and perform a more rigorous renormalization group analysis
in the second step. At time $t$ after the quench correlations are
limited by the light-cone. This gives rise to a time dependent mass
$r\left(t\right)=\gamma a/t^{2/z}$ in the propagator, where $a$
is a dimensionless coefficient. Scattering events caused by collisions
of excitations in regions of $t$-dependent size turn out to be highly
singular. A perturbation theory in $a$ that includes such scattering
events yields to leading order and for $t'\ll t$:
\begin{align}
G^{R}(k,t,t') & =G_{0}^{R}(k,t)\bigl[1 + \theta\log(t/t') + \ldots \bigr]\,,
\label{eq:6}
\end{align}
where the omitted terms are non-singular for $t'\rightarrow 0$ and 
\begin{equation}
\theta=-\frac{a\sin(\pi/z)}{\Gamma(2/z)}\,.\label{exponent of the retarded function}
\end{equation}
$G_{0}^{R}$ is the bare retarded Green's function given in the supplementary
section~\cite{Supplemental}. Exponentiation of the logarithm leads
to Eq.\eqref{eq:scaling}.

We now perform a momentum-shell RG approach to sum up these logarithms
in a controlled fashion and determine the exponent $\theta$ . In
full analogy to the equilibrium case we integrate out states in a
shell with momenta $\Lambda/b<k<\Lambda$ with $b>1$ and rescale
fields, momenta and time variables. The small parameter controlling
the calculation is the deviation from the upper critical dimension
$\epsilon=4-d-z$. The mass $\delta r_{i}$ in the initial Hamiltonian
is a strongly relevant perturbation and rapidly flows to large values.
The non-equilibrium dynamics of the system is therefore governed by
the deep-quench fixed point $\left(\hat{u}^{*},\delta r_{i}^{*},\delta r_{f}^{*}\right)=\left(\hat{u}^{*},\infty,0\right)$.
Here $\hat{u}^{*}=\frac{c_{z}}{N+8}\epsilon$ is the equilibrium value
of the dimensionless coupling constant $\hat{u}=uK_{d}\Lambda^{-\epsilon}/\gamma^{z/2}$
with $K_{d}=\frac{\Gamma(d/2)}{2\pi^{d/2}(2\pi)^{d}}$ and coefficient
$c_{z}=\frac{4\sin(\pi z/2)}{z\left(2-z\right)\sin^{z/2}(\pi/z)}$.
The scaling dimension of $\delta r_{i}$ is relative to the fixed
point $\delta r_{i}^{*}=\infty$, i.e. $1/\delta r_{i}\propto b^{-\kappa/\nu}$
is a dangerously irrelevant variable at the deep quench fixed point. 

\begin{figure}[t!]
  \centering
  \includegraphics[width=.85\linewidth]{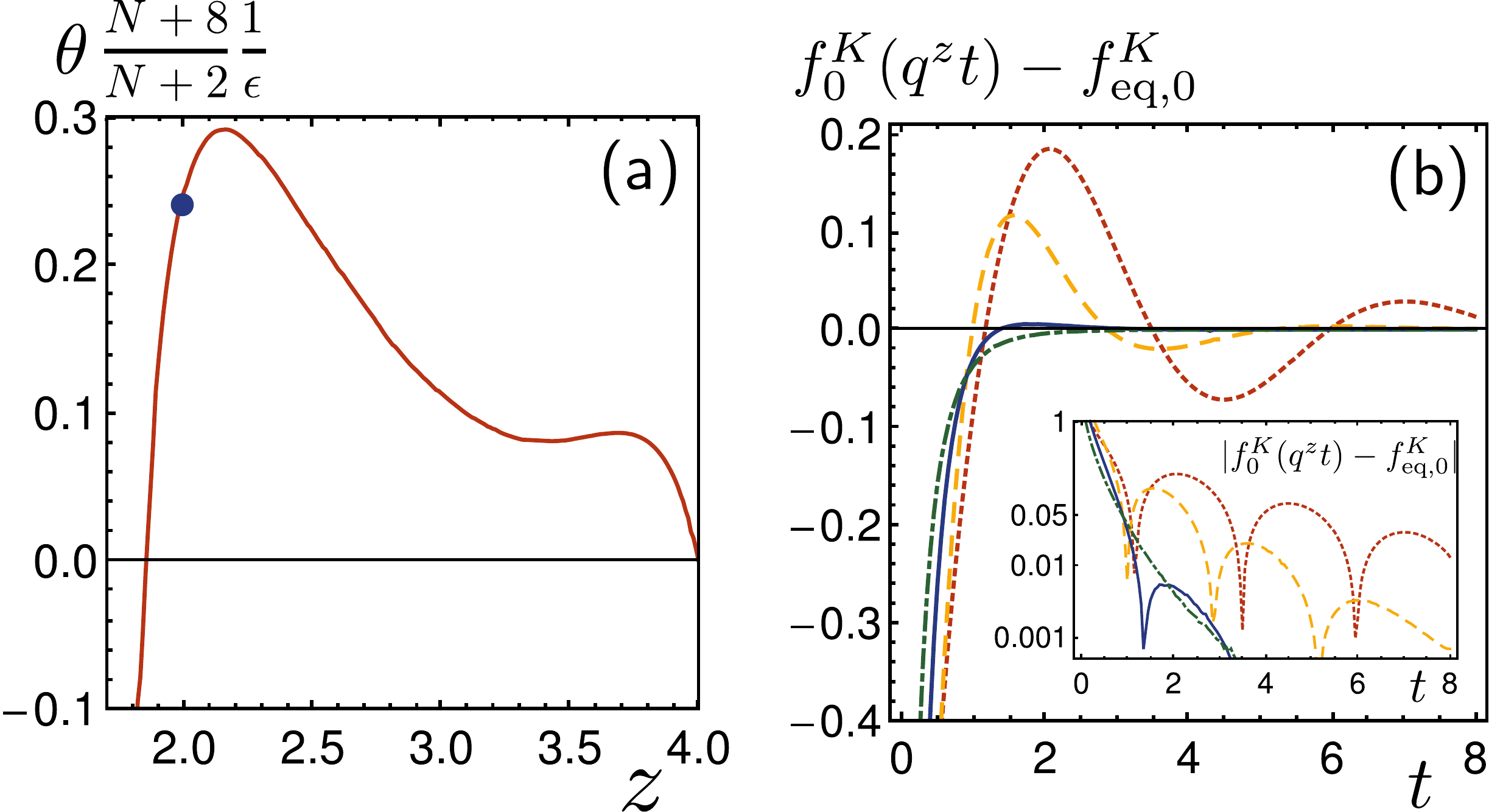}
  \caption{(Color online) (a) Prethermalization exponent $\theta$ as a function of dynamic critical exponent $z$. Plot shows $\mathcal{C}_z = \theta \frac{N+8}{N+2} \frac{1}{\epsilon}$, where $N$ is the number of components of $\boldsymbol{\varphi}$ and $\epsilon = 4 - d - z$. Blue dot indicates analytical result of Eq.~\eqref{eq:1}. (b) Free Keldysh scaling function $f^K_0(q^z t) - f_{\text{eq},0}$ after the quantum quench for different dynamic critical exponents $z = 1.2$ (red), $1.4$ (yellow), $2$ (blue), $2.5$ (green). Inset shows exponential decay of the envelope towards the equilibrium distribution, which becomes algebraic in presence of interactions. }
  \label{fig:3}
\end{figure}

We work with $\delta r_{i}>0$ corresponding to a quench out of the
unbroken phase and assume that $\theta$ is the same for the two paths
$A\rightarrow C$ and $B\rightarrow C$. For the mass renormalization after the quench follows
at one-loop 
\begin{equation}
r_{f}'\left(t\right)=b^{2}r_{f}\left(b^{z}t\right)+u\frac{N+2}{2}\int^{>}\frac{d^{d}k}{\left(2\pi\right)^{d}}iG_{0}^{K}\left(k,t,t\right)\,,\label{eq:2}
\end{equation}
where $>$ refers to momenta inside the shell. In equilibrium $r_{f}\left(t\right)$
is $t$-independent and we recover the usual one-loop result for the
mass renormalization. The quench mixes $r_{f}\left(t\right)$ at different
times during the flow. For a similar analysis of classical surface
criticality, see Ref.~\onlinecite{CordereyGriffin-AnnPhys-1981}.
We replace $\delta r_{i}$, that enters $G_{0}^{K}$, and $\hat{u}$
by their deep-quench fixed-point values. From Eq.~\eqref{eq:2} we
then obtain a differential equation for the corresponding time-dependent
fixed-point mass $r_{f}^{*}\left(t\right)$: 
\begin{equation}
2r_{f}^{*}+zt\frac{dr_{f}^{\ast}}{dt}+\frac{(N+2)\hat{u}^{*}\Lambda^{2}}{2}f_{0}^{K}\bigl(\Lambda^{z}t/\gamma^{z/2},1\bigr)=0\,.\label{eq:dgl}
\end{equation}
The scaling function $f_{0}^{K}$ characterizes $G_{0}^{K}$ according
to Eq.~\eqref{eq:scaling}. The solution of Eq.~\eqref{eq:dgl}
is 
\begin{equation}
r_{f}^{*}\left(t\right)=\frac{\gamma a}{t^{2/z}}-\frac{(N+2)\hat{u}^{*}\Lambda^{2}}{2zt^{2/z}}\int^{t}dt'f_{0}^{K}\left(\frac{\Lambda^{z}t'}{\gamma^{z/2}},1\right)t'^{\frac{2-z}{z}}\,,\label{eq:FPsol}
\end{equation}
where $a$ denotes the integration constant of Eq.~\eqref{eq:dgl}.
We find $f_{0}^{K}\left(\Lambda^{z}t/\gamma^{z/2}\rightarrow\infty,1\right)\rightarrow f_{eq,0}^{K}$,
where $f_{eq,0}^{K}$ describes the equal-time Keldysh function in equilibrium
after the quench. For a perturbative RG analysis a long range decay
of the mass parameter cannot emerge. We can therefore fix the integration
constant $a$ from the condition that $r_{f}^{*}(t)$ rapidly approaches
its equilibrium value, i.e. that $\delta r_{f}^{*}\left(t\right)=r_{f}^{*}\left(t\right)-r_{{\rm eq}}^{*}\rightarrow0$
for $t\gg\gamma^{z/2}\Lambda^{-z}$: 
\begin{equation}
a=\frac{(N+2)\hat{u}^{*}}{2z}\int_{0}^{\infty}dx\left(f_{0}^{K}\left(x,1\right)-f_{eq}^{K}\right)x^{\frac{2-z}{z}}\,.\label{eq:condition1}
\end{equation}
The derivation of the free non-equilibrium Keldysh function $G_{0}^{K}$
and thus of $f_{0}^{K}\left(x,1\right)$ is given in the supplementary
section~\cite{Supplemental}. Once we determine the coefficient $a$,
the exponent $\theta$ follows from Eq.~\eqref{exponent of the retarded function}.
For an ohmic bath with $\alpha=1$, i.e. $z=2$, we find analytically
$a_{z=2}=-\frac{N+2}{N+8}\frac{\epsilon}{4}$, which yields with Eq.~\eqref{exponent of the retarded function}
the exponent~\cite{Supplemental} 
\begin{equation}
\label{eq:1}
\theta_{z=2}=\frac{N+2}{N+8}\frac{1}{4}\epsilon>0.
\end{equation}
For a bath with colored noise, we determine the exponent numerically.
Our results for $\mathcal{C}_z = \theta \frac{N+8}{N+2} \frac{1}{\epsilon}$ are shown in Fig.~\ref{fig:3}(a). We find a maximal
value for $\mathcal{C}_z$ (and thus $\theta$) in the slightly sub-ohmic regime, while $\theta(z \rightarrow 4)\rightarrow 0$ since $\epsilon>0$ requires at least $z<4$. For $z<2$ the exponent decreases and changes sign for $z \approx 1.8$. From Eq.\eqref{eq:condition1} follows that the coefficient $a$ and thus $\theta$ can only change
sign if equal-time correlations decay non-monotonically. In Fig.~\ref{fig:3}(b) we show the scaling function $f_{0}^{K}\left(x,1\right)$ which proves that this is indeed the case for a super-ohmic bath. Note, in our analysis the limit $z\rightarrow1$ does not correspond to the closed system with ballistic time evolution as we always consider the limit of bath-dominated dynamics.

For $z=2$ the value of ${\cal C}_{z=2}$ turns out to be the same
as for a classical phase transition~\cite{JanssenSchaubSchmittmann-1989, 1742-5468-2012-01-P01014}. Idenical coefficients
for classical and quantum phase transitions might suggest that quantum
effects are not important for the quench dynamics. However, considering
generic values of $z$ the exponents (for given $\epsilon$) of a
classical and quantum quench are clearly distinct, demonstrating the
quantum quench dynamics is in a different universality class as the
classical one. 

Let us discuss the physical implications of these results: i) \emph{collapse
of the correlation length:} We compare the correlation length prior
to the quench $\xi_{i}\propto\delta r_{i}^{-\nu}$ with its value
at the crossover between the prethermalized regime and equilibration
$\xi\left(t^{*}\right)\propto\delta r_{i}^{-\nu/\kappa}$. $\theta>0$
implies with Eq.\eqref{eq:phi_of_t} that $\kappa>1$, such that $\xi\left(t^{*}\right)<\xi_{i}$ for small $\delta r_{i}$. Right after the quench the system
falls out of equilibrium and breaks up into many small uncorrelated regions. The
correlation length collapses and does not reach its pre-quench value
during prethermalization. It takes until after the time scale $t^{*}$
that the system recovers its initial correlations (see spin configurations in Fig.~\ref{fig:1}(c)). ii) \emph{order
parameter dynamics:} From Eq.\eqref{eq:phi_of_t} follows for $\theta>0$
that the order parameter grows as function of time. The physical explanation
for this behavior follows from our discussion of the path $A\rightarrow C$. 
$\theta>0$ leads to a slowing down of the temporal decay of local
correlations. On the other hand, the size of correlated regions increases
according to the light-cone scale $\xi\left(t\right).$ The order
parameter grows because of the coarsening that takes place at intermediate
time scales where the growth in $\xi\left(t\right)$ outweighs the
decay of correlations. Thus, the growth of the order parameter $\propto t^{\theta}$
is caused by the recovery of locally ordered regions after the collapse
of the correlation length. The long-time, quasi-adiabatic order-parameter
dynamics $\left\langle \varphi_{a}\left(t\right)\right\rangle \propto\xi\left(t\right)^{-\beta/\nu}$only
sets in when initial correlations are recovered. iii) \emph{equal
time correlations:} a straightforward extension of our RG analysis
to the scaling function $f^{K}$ in Eq.~\eqref{eq:scaling} yields,
instead of the exponential decay shown in Fig.~\ref{fig:3}(b), a power law decay
$f^{K}\left(x,1\right)=f_{{\rm eq}}^{K}-\frac{2\theta}{c_{z}\sin\frac{\pi}{z}}x^{-2/z}$ with universal coefficient proportional to $\theta$. 
iv)$ $\emph{ the regime with $\theta<0$:} In this
case no coarsening growth of the order parameter occurs, yet its
decay is slowed down if compared to the quasi-adiabatic regime. In
addition, the correlation length recovers before the crossover time
$t^{*}$ is reached. v) \emph{duration of prethermalization:} Since
the crossover time $t^{*}$ diverges for weak quenches, an almost
critical system, subject to a sudden change of its parameters, undergoes
universal out-of-equilibrium dynamics for arbitrarily long periods
of time.

In conclusion, we determined universal behavior that governs quantum
critical prethermalization. The intermediate time dynamics of a system
that is suddenly moved to a nearby QCP is characterized by a new exponent $\theta$. Owed to the quench,
the system instantly falls out of equilibrium and breaks up into small
correlated regions. The quantum critical prethermalization describes
the recovery after this collapse and extends over long times, depending
on the initial distance from the critical point. A quench close to
a quantum critical point opens the possibility to quantitatively analyze
the universal far-from-equilibrium dynamics of a many body system
and to manipulate the crossover between prethermalization and thermalization
regimes.

The Young Investigator Group of P.P.O. received financial support
from the ``Concept for the Future'' of the KIT within the framework
of the German Excellence Initiative.

%

\pagebreak


\section*{Supplementary material}

# Supplementary Material to "Universal post-quench prethermalization at a quantum critical point"

Pia Gagel,[1] Peter P. Orth,[1] and Jörg Schmalian[1, 2]

[1]*Institute for Theory of Condensed Matter, Karlsruhe Institute of Technology (KIT), 76131 Karlsruhe, Germany*
[2]*Institute for Solid State Physics, Karlsruhe Institute of Technology (KIT), 76021 Karlsruhe, Germany*

## BARE RESPONSE AND CORRELATION FUNCTIONS

In order to develop a perturbative renormalization group approach we first need to determine the bare response and correlation functions. In distinction to the equilibrium case, the non-interacting problem far from equilibrium is nontrivial in its own right as we have to determine the bare propagators for our quench-protocol.

For the non-interacting problem it is straightforward to solve the Heisenberg equation of motion of the field operator $\boldsymbol{\varphi}(k,t)$. We find for the Laplace transform

$$\boldsymbol{\varphi}(k,\omega) = \int_0^\infty \boldsymbol{\varphi}(k,t) e^{i(\omega+i0^+)t} dt$$
$$= \mathbf{F}(k,\omega) G_{0,\text{eq}}^R(k,\omega), \tag{1}$$

with force-operator

$$\boldsymbol{F}(k,\omega) = \mathbf{h}(-k,\omega) + \boldsymbol{\pi}_0(k) - i\omega\boldsymbol{\varphi}_0(k) + \boldsymbol{\xi}(\omega). \tag{2}$$

$\mathbf{h}(k,\omega)$ is a field conjugate to $\boldsymbol{\varphi}(k,\omega)$ that has been added to the Hamiltonian. The $t=0$ initial values $\boldsymbol{\pi}_0(k)$ and $\boldsymbol{\varphi}_0(k)$ are determined by the pre-quench dynamics. Finally,

$$\boldsymbol{\xi}(t) = -\sum_l \left( c_l \boldsymbol{X}_{0l} \cos(\Omega_l t) + \frac{c_l \boldsymbol{P}_{0l}}{\Omega_l} \sin(\Omega_l t) \right) \tag{3}$$

is the operator of the bath-induced force, where $\boldsymbol{X}_{0l} = \boldsymbol{X}_l(t=0)$ and $\boldsymbol{P}_{0l} = \boldsymbol{P}_l(t=0)$ for the oscillator mode $l$ with frequency $\Omega_l$.

$$G_{0,\text{eq}}^R(k,\omega) = \frac{1}{r_0 + k^2 - \omega^2 + \eta(\omega)}. \tag{4}$$

is the bare retarded Green's function without quench, corresponding to equilibrium of the post quench Hamiltonian.

From this solution we now determine the bare retarded are Keldysh Green's functions

$$G_0^R(k,t,t') = -i\theta(t-t') \left\langle [\varphi_a(k,t), \varphi_a(-k,t')]_- \right\rangle$$
$$G_0^K(k,t,t') = -i \left\langle [\varphi_a(k,t), \varphi_a(-k,t')]_+ \right\rangle, \tag{5}$$

where $[A,B]_\pm = AB \pm BA$. It is convenient to use the double Laplace transform

$$G_0^{R/K}(k,\omega,\omega') = \int_0^\infty dt \int_0^\infty dt' G_0^{R/K}(t,t') e^{i(\omega_+ t + \omega'_+ t')} \tag{6}$$

where we use the notation $\omega_\pm = \omega \pm i0^+$.

For the retarded Green's function we use

$$G_0^R(k,\omega,\omega') = \frac{\delta \langle \varphi_a(k,\omega) \rangle}{\delta h_a(-k,\omega')} \tag{7}$$

and take the expectation value of Eq. (2)

$$\langle \varphi_a(k,\omega) \rangle = h_a(-k,\omega) G_{0,\text{eq}}^R(k,\omega). \tag{8}$$

Using $\delta h_a(k,t)/\delta h_a(k,t') = \delta(t-t')$ it follows

$$\frac{\delta h_a(k,\omega)}{\delta h_a(k,\omega')} = \frac{i}{\omega+\omega'+i0^+}. \tag{9}$$

This yields the result:

$$G_0^R\left(k,\omega,\omega'\right) = \frac{-iG_{0,\text{eq}}^R\left(k,\omega\right)}{\omega+\omega'+i0^+}. \tag{10}$$

Using this result, it follows from simple contour integration that, despite the quench, the bare retarded Green's functions $G_0^R$ on the round trip contour only depends on the time difference $t-t'$ between two events. For the Fourier transform follows the same result as in equilibrium:

$$G_0^R\left(k,\omega\right) = G_{0,\text{eq}}^R\left(k,\omega\right). \tag{11}$$

The information about the quench is contained in the Keldysh function $G_0^K$. Inserting the solution of the Heisenberg equation of motion we obtain

$$G_0^K\left(k,\omega,\omega'\right) = M\left(k,\omega,\omega'\right)G_0^R\left(k,\omega\right)G_0^R\left(k,\omega'\right), \tag{12}$$

where we introduced the memory function

$$M\left(k,\omega,\omega'\right) = -i\left\langle\left[F_a\left(k,\omega\right),F_a\left(-k,\omega'\right)\right]_+\right\rangle. \tag{13}$$

In the time domain, Eq. (12) is given as

$$G_0^K\left(k,t,t'\right) = \int_0^\infty dsds' M\left(k,s,s'\right) \\ \times G_0^R\left(k,t-s\right)G_0^R\left(k,t'-s'\right). \tag{14}$$

This is the out-of-equilibrium version of the fluctuation-dissipation theorem for our quench protocol.

The remaining task is to determine the force-force correlation function that determines $M\left(k,\omega,\omega'\right)$. Here, the bath induced long-time correlations between pre- and post-quench dynamics, indicated in the Keldysh contour, correspond to mixed terms like $-i\left\langle\left[\pi_a\left(k,\omega\right),\xi_a\left(-k,\omega'\right)\right]_+\right\rangle$ in the memory function. Of the numerous terms that emerge many have contributions that diverge in the limit $\omega_c \to \infty$, with bath cut off $\omega_c$. A straightforward but rather tedious analysis yields that all divergencies cancel and leads to our result for the the memory function:

$$M\left(k,\omega,\omega'\right) = -i\frac{G_{0,i}^K\left(k,\omega\right)+G_{0,i}^K\left(k,\omega'\right)}{\omega+\omega'+i0^+} \\ \times G_{0,i}^R\left(k,\omega\right)^{-1}G_{0,i}^R\left(k,\omega'\right)^{-1}. \tag{15}$$

Here, the index $i$ refers to the retarded Green's function $G_{0,i}^R\left(k,\omega\right)$ and the Laplace transform of the Keldysh function $G_{0,i}^K\left(k,t-t'\right)$ of a system in equilibrium prior to the quench. Without quench $G_{0,i}^{R(K)} = G_0^{R(K)}$ and one recovers after a few steps that $G_0^K\left(k,t,t'\right)$ of Eq. (14) takes its equilibrium value, obeying the fluctuation-dissipation theorem. In the classical limit we obtain the results of Ref. 1 for an ohmic bath and Ref. 2 for colored noise.

## SCALING LIMIT AND EVALUATION OF THE EXPONENT

From the RG-analysis in the main paper follows for the critical exponent

$$\theta = -\frac{\sin\left(\frac{\pi}{z}\right)}{\Gamma\left(\frac{2}{z}\right)}\frac{u^*\left(N+2\right)\Lambda^{-\epsilon}K_d}{2z\gamma^{z/2}}I_0 \tag{16}$$

with

$$I_0 = \frac{q^{4-z}}{\gamma^{\frac{2-z}{2}}}\int_0^\infty dt\left(iG_0^K\left(k,t,t\right)-iG_{0\text{eq}}^K\left(k\right)\right)t^{\frac{2-z}{2}} \tag{17}$$

and $u^*$ the fixed point value of the interaction. For the Keldysh function we use the result:

$$iG_0^K\left(k,t,t\right) = \int_0^\infty ds\int_0^\infty ds' M\left(s,s',\omega_0^2\right)G_0^R\left(k,t-s\right)G_0^R\left(k,t-s'\right), \tag{18}$$

where the argument $\omega_0^2 = \delta r_i + k^2$ indicates that $M$ is obtained from equilibrium Green's functions prior to the quench. $\delta r_i$ is a measure for the distance from the critical point for $t < 0$. The equilibrium Keldysh function after the quench is

$$i G_{0\text{eq}}^K (k) = \int_0^\infty ds \int_0^\infty ds' M \left(s, s', k^2\right) G_0^R \left(k, t-s\right) G_0^R \left(k, t-s'\right), \tag{19}$$

If we introduce

$$\delta M \left(s, s'\right) = M \left(s, s', \omega_0^2\right) - M \left(s, s', q^2\right), \tag{20}$$

it follows

$$I_0 = \frac{q^{4-z}}{\gamma^{\frac{2-z}{2}}} \int_0^\infty ds \int_0^\infty ds' \delta M \left(s, s'\right) \int_0^\infty dt\, t^{\frac{2-z}{z}} G_0^R \left(k, t-s\right) G_0^R \left(k, t-s'\right). \tag{21}$$

Instead of integrations over time, we can easily rewrite this in terms of the frequency dependent memory and retarded Green's functions:

$$I_0 = \frac{-\Gamma \left(\frac{2}{z}\right)}{i^{2/z}} \frac{q^{4-z}}{\gamma^{\frac{2-z}{2}}} \int_{-\infty}^\infty \frac{d\omega d\omega'}{\pi^2} \frac{\text{Im} G_0^R \left(k, \omega\right) \text{Im} G_0^R \left(k, \omega'\right) \delta M \left(\omega, \omega'\right)}{\left(\omega + \omega' - i 0^+\right)^{2/z}}, . \tag{22}$$

The evaluation of the integrals is straighforward. For $z = 2$ it can be done analytically and we find

$$I_0 \left(z = 2\right) = -\frac{1}{\pi}$$

of Eq. (17). Performing the limit $z \to 2$ in the coefficient in Eq. (16) in front of $I_0$, we obtain the result for the exponent given in the main paper:

$$\theta \left(z = 2\right) = \frac{1}{4} \frac{N+2}{N+8} \epsilon. \tag{23}$$

For $z > 2$ the integral Eq. (22) is performed numerically.

-------------------



\end{document}